\documentclass{article}
\usepackage{spconf}
\usepackage{dsfont}
\usepackage{mathdots}
\usepackage{color}
\usepackage{graphicx}
\usepackage{amsmath}
\usepackage{epstopdf}
\usepackage{amsthm}
\usepackage{cite}
\usepackage{amssymb}
\usepackage{epsfig}
\usepackage{amsfonts}
\usepackage[]{algorithmic}
\usepackage[]{algorithm2e}
\usepackage{listings}
\usepackage[keeplastbox]{flushend}
\usepackage{fancybox}
\usepackage{epsfig}
\usepackage{mathtools}

\newcommand{\Np}{N_{\mathrm{p}}}
\newcommand{\imj}{\mathsf{j}}
\usepackage{amsmath}

\title{Low-rank mmWave MIMO channel estimation in one-bit receivers}
\name{Nitin Jonathan Myers, Kayla N. Tran and Robert W. Heath Jr. \thanks{This research was supported in part by the National Science Foundation under Grant numbers NSF-CNS-1731658 and NSF-CNS-1702800. K. N. Tran participated in this research through the Graduates Linked with Undergraduates in Engineering program at The University of Texas at Austin.}}
\address{Department of Electrical and Computer Engineering, The University of Texas at Austin.
  \\{email :\small\tt \{nitinjmyers,kaylatran,rheath\}@utexas.edu }}

\begin{document}
\ninept
\maketitle
\begin{abstract}
Receivers with one-bit analog-to-digital converters (ADCs) are promising for high bandwidth millimeter wave (mmWave) systems as they consume less power than their full resolution counterparts. The extreme quantization in one-bit receivers and the use of large antenna arrays at mmWave make channel estimation challenging. In this paper, we develop channel estimation algorithms that exploit the low-rank property of mmWave channels. We also propose a novel training solution that results in a low complexity implementation of our algorithms. Simulation results indicate that the proposed methods achieve better channel reconstruction than compressed sensing-based techniques that exploit sparsity of mmWave channels. 
\end{abstract}
%
\vspace{-2mm}
\section{Introduction}
The large bandwidths used in millimeter wave (mmWave) systems motivate the need for high speed analog-to-digital converters (ADCs) \cite{ranganmmwave}. Designing high resolution ADCs for large bandwidth systems, however, can be difficult under cost and power budget constraints \cite{heathoverview}. A possible approach to meet both these constraints is to reduce the resolution of the ADCs. In the extreme case, the ADC resolution can be as low as one-bit. One-bit receivers, i.e., receivers with one-bit ADCs for the in-phase and quadrature-phase components, are promising due to their low cost and low power consumption. Such receivers can only obtain the sign of the incoming signal due to coarse quantization at the ADCs.
\par Multiple-input multiple-output (MIMO) channel estimation in one-bit receivers is challenging due to quantized channel measurements \cite{one_bit_massive}. The use of large antenna arrays in mmWave MIMO systems further complicates the problem. Prior work has considered MIMO channel reconstruction with one-bit measurements \cite{junil_one_bit_chest,jianhua_asilomar,aldebaro_chest}. Existing methods are based on maximum likelihood estimation \cite{junil_one_bit_chest}, compressed sensing \cite{sp_lowres_p1,sp_lowres_p2,sp_lowres_p3,sp_lowres_p4,sp_lowres_p5,jianhua_asilomar}, and deep learning \cite{aldebaro_chest,deep_chest_2,deep_chest_3}. Compressed sensing-based methods exploit the sparse structure in channels, and deep learning-based techniques learn the channel structure using a data-driven approach. In this paper, we show that algorithms that exploit channel structure in forms other than sparsity may perform better. 
\par MmWave MIMO channels can be approximated as low-rank due to clustering in the propagation environment. The low-rank property was used in \cite{rappaport_low_rank} and \cite{joint_sp_rk_est} for channel estimation using an analog beamforming system with full resolution ADCs. To the best of our knowledge, low-rank MIMO channel estimation from one-bit measurements has not been studied. Prior work has developed theory and algorithms to recover low-rank matrices with real entries from one-bit measurements \cite{davenport_one_bit}. The direct application of the ideas in \cite{davenport_one_bit} to the channel estimation problem, however, is not straightforward due to complex valued channel matrices and hardware constraints. For example, sampling the entries of the low-rank channel matrix can result in low signal-to-noise (SNR) measurements under a per-antenna power constraint at the transmitter. In this paper, we solve these practical challenges and develop low-rank MIMO channel estimation algorithms for one-bit receivers. We also propose a training solution that allows acquiring a large number of distinct channel measurements without a substantial increase in the computational complexity of our algorithms. The algorithms and training proposed in this paper assume a narrowband MIMO channel. The narrowband assumption is simplistic and our solution can be extended to wideband systems using the one-bit tensor completion framework in \cite{one_bit_tensor}.
\par  \textbf{Notation}$:$ $\mathbf{A}$ is a matrix, $\mathbf{a}$ is a column vector and $a, A$ denote scalars. Using this notation $\mathbf{A}^T$ and $\mathbf{A}^{\ast} $ represent the transpose and conjugate transpose of $\mathbf{A}$. The real and imaginary parts of $\mathbf{A}$ are denoted by $\mathbf{A}^\mathrm{R}$ and $\mathbf{A}^\mathrm{I}$. We use $\mathrm{diag}\left(\mathbf{a}\right)$ to denote a diagonal matrix with entries  of $\mathbf{a}$ on its diagonal. The nuclear norm and the Frobenius norm of $\mathbf{A}$ are denoted by $\Vert \mathbf{A} \Vert_{\ast}$ and $\Vert \mathbf{A} \Vert_{\mathrm{F}}$~\cite{Linear_Alg_text}. The scalar $A_{k,\ell}$ denotes the entry of $\mathbf{A}$ in the $k^{\mathrm{th}}$ row and the ${\ell}^{\mathrm{th}}$ column. $\mathds{1}_{[\cdot]}$ is the indicator function and $\mathbf{I}$ denotes the identity matrix. The $\mathrm{sign}$ function is defined as $\mathrm{sign}(a)=1$ for $a\geq0$ and  $\mathrm{sign}(a)=-1$ for $a<0$. $\mathsf{j}=\sqrt{-1}$.
\vspace{-3mm}
\section{System model}
We consider a narrowband MIMO system with $N$ antennas at the transmitter (TX). For ease of notation, we assume that the number of antennas at the receiver (RX) is $N$. Each antenna at the RX is equipped with a pair of one-bit ADCs that output the sign of the in-phase and the quadrature-phase signals. The propagation environment between the TX and the RX is modeled by a narrowband channel matrix $\mathbf{H} \in \mathbb{C}^{N \times N}$. The one-bit quantization effect at the RX is modeled using the function $\mathcal{Q}_1(\cdot)$ defined as 
\begin{equation}
\label{eq:quantizer_defn}
\mathcal{Q}_1(\mathbf{A})=\mathrm{sign}(\mathbf{A}^{\mathrm{R}})+\mathsf{j} \, \mathrm{sign}(\mathbf{A}^{\mathrm{I}}).
\end{equation}
The $\mathrm{sign}(\cdot)$ operation is performed element-wise over the matrices $\mathbf{A}^{\mathrm{R}}$ and $\mathbf{A}^{\mathrm{I}}$ in \eqref{eq:quantizer_defn}. We define $\mathbf{v}\in \mathbb{C}^N$ as a complex additive white Gaussian noise (AWGN) vector with statistics $\mathbf{v}\sim \mathcal{N}_c(\mathbf{0},2\sigma^2\mathbf{I})$. The one-bit quantized vector received at the RX when the TX transmits a pilot vector $\mathbf{s}\in \mathbb{C}^{N}$ is given by
\begin{equation}
\label{eq:y_vect_defn}
\mathbf{y}=\mathcal{Q}_1(\mathbf{H}\mathbf{s}+\mathbf{v}).
\end{equation}
It can be observed from \eqref{eq:y_vect_defn} that the $N-$antenna RX acquires $2N$ bits of information for every transmission by the TX. 
\par Channel estimation in MIMO systems allows efficient data transmission and interference management. Estimating the mmWave MIMO channel with one-bit receivers, however, can be hard due to two reasons.  First, the dimension of a typical mmWave MIMO channel, i.e., $N \times N$, is larger than conventional lower frequency MIMO systems. Second, complex valued entries in $\mathbf{H}$ must be recovered from just sign-based channel measurements. In this case, algorithms that exploit special structure in mmWave channels can achieve better channel reconstruction than conventional techniques. Prior work has shown that typical mmWave channels can be approximated as low-rank, i.e., $\mathrm{rank}(\mathbf{H})\ll N$ \cite{rappaport_low_rank,joint_sp_rk_est}. The low-rank structure in mmWave MIMO channels allows applying ideas from one-bit matrix completion to channel estimation in one-bit receivers.
\section{Low-rank MIMO channel estimation \\ with one-bit measurements}
\par We explain the key ideas underlying our channel estimation techniques by considering $N_\mathrm{p}=N$ pilot transmissions. We define a training block as a matrix $\mathbf{S} \in \mathbb{C}^{N \times N}$ that is known to both the TX and the RX. The TX transmits the $n^{\mathrm{th}}$ column of $\mathbf{S}$ as the $n^{\mathrm{th}}$ pilot vector. We use $\mathbf{V}\in \mathbb{C}^{N \times N}$ to denote a noise matrix whose entries are IID and are distributed as $\mathcal{N}_\mathrm{c}(0, 2\sigma^2)$. The received block $\mathbf{Y}$ when the TX transmits $\mathbf{S}$ can be expressed as
\begin{equation}
\label{eq:Yblk_N}
\mathbf{Y}=\mathcal{Q}_1(\mathbf{H}\mathbf{S}+\mathbf{V}).
\end{equation}
Due to one-bit quantization, MIMO channel estimation from the $N^2$ measurements in \eqref{eq:Yblk_N} is an under-determined problem even in a noiseless setting. An infinite number of matrices in $\mathbb{C}^{N\times N}$ result in the same channel measurements as $\mathbf{H}$. Most of the matrices that satisfy \eqref{eq:Yblk_N}, however, may not have a low-rank. Such matrices are less likely to be mmWave channels. In this section, we develop optimization algorithms to estimate a matrix that has a low-rank and is faithful to the received one-bit channel measurements.
\subsection{Low-rank constraint and the log-likelihood function}
\par The optimization algorithms developed in this section solve for the transformed channel instead of the original channel. We explain how such a transformation reduces the complexity of our algorithms in Sec.~\ref{sec:complexity}. The transformed channel, called the pseudo-channel, is defined as
\begin{equation}
\label{eq:defn_G}
\mathbf{G}=\mathbf{H S}.
\end{equation}
As $\mathrm{rank}(\mathbf{HS})\leq \mathrm{rank}(\mathbf{H})$ and $\mathbf{H}$ is assumed to be a low-rank matrix, $\mathbf{G}$ in \eqref{eq:defn_G} has a low-rank. In this paper, the low-rank matrix $\mathbf{G}$ is first estimated from the channel measurements. Then, the transformation in \eqref{eq:defn_G} is inverted to estimate $\mathbf{H}$ from $\mathbf{G}$. We define $\mathbf{X} \in \mathbb{C}^{N \times N}$ as the optimization variable corresponding to $\mathbf{G}$. The low-rank constraint on $\mathbf{X}$, i.e., $\mathrm{rank}(\mathbf{X})=r$ for some $r\ll N$, however, is non-convex. A common approach to solve this problem is to relax the low-rank constraint into a nuclear norm constraint \cite{candes_MC}, i.e., 
\begin{equation}
\label{eq:nuc_norm_const}
\Vert \mathbf{X} \Vert _{\ast} \leq \beta, 
\end{equation}
for some constant $\beta$. For low-rank channels, the nuclear norm constraint results in better channel reconstruction than the $\ell_2$-norm constraint used in \cite{junil_one_bit_chest} . In practice, $\beta$ can be determined from the channel statistics and the training $\mathbf{S}$ using $\Vert \mathbf{HS} \Vert _{\ast} \leq \beta$. The constraint in \eqref{eq:nuc_norm_const} represents a nuclear norm ball which is a convex set~\cite{boyd2004convex}. 
\par Now, we derive the log-likelihood function that quantifies how well a matrix is consistent with the received channel measurements. It can be observed from \eqref{eq:Yblk_N} and \eqref{eq:defn_G} that 
\begin{equation}
\label{eq:Y_blk_G_N}
\mathbf{Y}=\mathcal{Q}_1(\mathbf{G+\mathbf{V}}).
\end{equation} 
The optimization variable $\mathbf{X}$ is expected to satisfy $\mathbf{Y}=\mathcal{Q}_1(\mathbf{X+\mathbf{N}})$, for some noise matrix $\mathbf{N}$ that has the same statistics as $\mathbf{V}$. Note that the in-phase one-bit measurement $Y^{\mathrm{R}}_{k,\ell}$ is $1$ when $G^{\mathrm{R}}_{k,\ell}+V^{\mathrm{R}}_{k,\ell}>0$ and is $-1$ when $G^{\mathrm{R}}_{k,\ell}+V^{\mathrm{R}}_{k,\ell}<0$. The stochastic nature of $\mathbf{V}$ defines a probability distribution on the received one-bit measurements. We use $\Phi(\cdot)$ to denote the cumulative distribution function of the standard normal random variable. As $\mathbf{Y}^{\mathrm{R}}=\mathcal{Q}_1(\mathbf{G^{\mathrm{R}}+\mathbf{V}^{\mathrm{R}}})$, the log-likelihood corresponding to $\mathbf{X}^{\mathrm{R}}$ can be expressed as \cite{davenport_one_bit}
\begin{align}
\nonumber
\mathcal{L}_{\mathbf{Y}^{\mathrm{R}}}(\mathbf{X}^{\mathrm{R}})&=\sum^{N}_{k=1}\sum^{N}_{\ell=1} \big[\mathds{1}_{[{Y}_{k,\ell}^{\mathrm{R}}=1]}\mathrm{log}\left( \Phi(X^{\mathrm{R}}_{k,\ell}/\sigma)\right)\\
\label{eq:likelihood_Real}
&+\mathds{1}_{[{Y}_{k,\ell}^{\mathrm{R}}=-1]}\mathrm{log}\left( 1-\Phi(X^{\mathrm{R}}_{k,\ell}/\sigma)\right)\big].
\end{align}
We define the log-likelihood corresponding to $\mathbf{X}^{\mathrm{I}}$ as $\mathcal{L}_{\mathbf{Y}^{\mathrm{I}}}(\mathbf{X}^{\mathrm{I}})$. The function $\mathcal{L}_{\mathbf{Y}^{\mathrm{I}}}(\mathbf{X}^{\mathrm{I}})$ is derived by replacing the superscript $\mathrm{R}$ in \eqref{eq:likelihood_Real} to $\mathrm{I}$. As $\mathbf{V}^{\mathrm{R}}$ and  $\mathbf{V}^{\mathrm{I}}$ are independent, the log-likelihood of $\mathbf{X}$ can be expressed as the sum of the log-likelihoods corresponding to the in-phase and quadrature phase components of $\mathbf{X}$, i.e., 
\begin{equation}
\label{eq:likelihood}
\mathcal{L}_{\mathbf{Y}}(\mathbf{X})=\mathcal{L}_{\mathbf{Y}^{\mathrm{R}}}(\mathbf{X}^{\mathrm{R}})+\mathcal{L}_{\mathbf{Y}^{\mathrm{I}}}(\mathbf{X}^{\mathrm{I}}).
\end{equation}
It can be verified that both $\mathcal{L}_{\mathbf{Y}^{\mathrm{R}}}(\mathbf{X}^{\mathrm{R}})$ and $\mathcal{L}_{\mathbf{Y}^{\mathrm{I}}}(\mathbf{X}^{\mathrm{I}})$ are concave in $\mathbf{X}$. By the property that the sum of two concave functions is concave, $\mathcal{L}_{\mathbf{Y}}(\mathbf{X})$ is concave in $\mathbf{X}$.
\subsection{Optimization algorithms for pseudo-channel estimation}\label{sec:algorithms}
\par A possible approach to estimate the pseudo-channel $\mathbf{G}$ is to maximize the log-likelihood function $\mathcal{L}_{\mathbf{Y}}(\mathbf{X})$ subject to the low-rank constraint on $\mathbf{X}$. With the nuclear norm relaxation of the low-rank constraint, the optimization problem can be formulated as \cite{davenport_one_bit}
\begin{equation}
\label{eq:generic_optim}
\hat{\mathbf{G}}=\mathrm{arg\,max}\,\mathcal{L}_{\mathbf{Y}}(\mathbf{X})\;\mathrm{subject\,\,to} \,\Vert \mathbf{X} \Vert _{\ast} \leq \beta.
\end{equation}
To solve \eqref{eq:generic_optim}, we develop two iterative optimization algorithms that are based on projected gradient ascent (PGA) \cite{boyd2004convex} and Franke-Wolfe \cite{fwm_jaggi} techniques. The PGA method was studied in \cite{davenport_one_bit} for one-bit recovery of matrices in $\mathbb{R}^{N \times N}$. We use $\mathbf{X}_t \in \mathbb{C}^{N\times N}$ to denote the optimization variable corresponding to $\mathbf{X}$ in the $t^{\mathrm{th}}$ iteration of an algorithm. Furthermore, we define $\nabla \mathcal{L}_{\mathbf{Y}}(\mathbf{X}_t) \in \mathbb{C}^{N \times N}$ as the gradient of $\mathcal{L}_{\mathbf{Y}}(\mathbf{X})$ at $\mathbf{X}= \mathbf{X}_t$. Specifically, the $(k,\ell)^{\mathrm{th}}$ entry of $\nabla \mathcal{L}_{\mathbf{Y}}(\mathbf{X}_t)$ is $\partial \mathcal{L}_{\mathbf{Y}^{\mathrm{R}}}(\mathbf{X}_t^{\mathrm{R}})/ \partial X^{\mathrm{R}}_{k,\ell}+\mathrm{j}\,\partial \mathcal{L}_{\mathbf{Y}^{\mathrm{I}}}(\mathbf{X}_t^{\mathrm{I}})/ \partial X^{\mathrm{I}}_{k,\ell}$. The concave nature of the objective function and the convex constraint set in \eqref{eq:generic_optim} allow our algorithms to converge to a global optimum.
\par We now explain PGA-based estimation of $\mathbf{G}$ from $\mathbf{Y}$. For a step size of $\eta$, the ascent step in PGA shifts $\mathbf{X}_t$ by $ \eta \nabla \mathcal{L}_{\mathbf{Y}}(\mathbf{X}_t)$. The matrix obtained after shifting $\mathbf{X}_t$ is defined as $\mathbf{Z}_{t+1}$. It is important to note that $\mathbf{Z}_{t+1}$ may not belong to the constraint set, i.e., $\Vert\mathbf{X} \Vert _{\ast} \leq \beta$, even when $\mathbf{X}_t$ satisfies the constraint. The projection step in PGA finds a matrix within the constraint set that is closest to $\mathbf{Z}_{t+1}$. The projection, defined as $\mathbf{X}_{t+1}$, is derived using the singular value decomposition (SVD) of $\mathbf{Z}_{t+1}$ and a simplex projection \cite{nuclear_projection}. The PGA algorithm to estimate $\mathbf{G}$ is summarized in Algorithm \ref{alg:pga}. It can be noticed from \eqref{eq:likelihood_Real} and \eqref{eq:likelihood} that the complexity of a gradient step, i.e., computing $\nabla \mathcal{L}_{\mathbf{Y}}(\mathbf{X})$, is $\mathcal{O}(N^2)$. The complexity of the SVD step in PGA, however, is $\mathcal{O}(N^3)$. Therefore, every iteration of the PGA algorithm has a complexity of $\mathcal{O}(N^3)$.
\begin{algorithm}
\label{algo:PGA}
\caption{Projected gradient ascent method to estimate $\mathbf{G}$}
\label{alg:pga}
\begin{algorithmic}
\FOR{$t=1$ to $T_{\mathrm{max}}$}
\STATE $\mathbf{Z}_{t+1}=\mathbf{X}_{t}+\eta \nabla \mathcal{L}_{\mathbf{Y}}(\mathbf{X}_t)$
\STATE  Compute the SVD: $\mathbf{Z}_{t+1}=\mathbf{U}_{t+1}\mathrm{diag}(\mathbf{d}_{t+1})\mathbf{V}^{\ast}_{t+1}$
\STATE  $\boldsymbol{\pi}_{t+1}\leftarrow$ Projection of $\mathbf{d}_{t+1}$ on $\{\mathbf{d}:\mathbf{1}^T\mathbf{d}=\beta, \mathbf{d}\geq0\}$
\STATE  $\mathbf{X}_{t+1}=\mathbf{U}_{t+1}\mathrm{diag}(\boldsymbol{\pi}_{t+1})\mathbf{V}^{\ast}_{t+1}$ 
\STATE Stop if $0<\mathcal{L}_{\mathbf{Y}}(\mathbf{X}_{t+1})-\mathcal{L}_{\mathbf{Y}}(\mathbf{X}_t)<\epsilon |\mathcal{L}_{\mathbf{Y}}(\mathbf{X}_t)|$
\ENDFOR
\STATE $\hat{\mathbf{G}}=\mathbf{X}_{t+1}$
\vspace{1mm}
\end{algorithmic}
\end{algorithm}
\par The Franke-Wolfe method maximizes a linear approximation of the objective, i.e., $\mathcal{L}_{\mathbf{Y}}(\mathbf{X})$, in every iteration \cite{fwm_jaggi}. As the linear approximation is determined by the gradient, this method selects a matrix $\mathbf{D}_t$ within $\Vert\mathbf{X} \Vert _{\ast} \leq \beta$ that maximizes the inner product $\langle\mathbf{D}_t, \nabla \mathcal{L}_{\mathbf{Y}}(\mathbf{X}_t) \rangle$. The matrix $\mathbf{D}_t$ is simply the rank-one approximation of the gradient $\nabla \mathcal{L}_{\mathbf{Y}}(\mathbf{X}_t)$ \cite{fwm}. For a step size of $\gamma_t$, the optimization variable $\mathbf{X}_t$ is incremented by $\gamma_t (\mathbf{D}_t-\mathbf{X}_t)$ to obtain $\mathbf{X}_{t+1}$. A summary of the Franke-Wolfe technique to estimate $\mathbf{G}$ is given in Algorithm \ref{alg:fwm}. To achieve a low complexity implementation of Algorithm \ref{alg:fwm}, we use the power method to compute the rank one-approximation of $\nabla \mathcal{L}_{\mathbf{Y}}(\mathbf{X}_t)$. Each iteration of the power method requires multiplying an $N \times N$ matrix with an $N \times 1$ vector. With the power method-based implementation, the complexity of a single iteration of the Franke-Wolfe method is $\mathcal{O}(N^2)$ which is lower than that of the PGA algorithm.
\vspace{-3mm}
\begin{algorithm}
\label{algo:FW}
\caption{Franke-Wolfe method to estimate $\mathbf{G}$}
\label{alg:fwm}
\begin{algorithmic}
\FOR{$t=1$ to $T_{\mathrm{max}}$}
\STATE $\mathbf{D}_{t}\leftarrow$ Rank $1$ approx. of $\nabla \mathcal{L}_{\mathbf{Y}}(\mathbf{X}_t)$ by power method
\STATE  $\gamma_t \leftarrow 2/(t+2)$
\STATE  $\mathbf{X}_{t+1}=\mathbf{X}_{t}+\gamma_t(\mathbf{D}_t-\mathbf{X}_{t})$
\STATE Stop if $0<\mathcal{L}_{\mathbf{Y}}(\mathbf{X}_{t+1})-\mathcal{L}_{\mathbf{Y}}(\mathbf{X}_t)<\epsilon |\mathcal{L}_{\mathbf{Y}}(\mathbf{X}_t)|$
\ENDFOR
\STATE $\hat{\mathbf{G}}=\mathbf{X}_{t+1}$
\vspace{1mm}
\end{algorithmic}
\end{algorithm}
\vspace{-6mm}
\subsection{Channel estimation and its complexity} \label{sec:complexity}
\par In this paper, we consider a unitary training block $\mathbf{S}$ so that the transformation between the pseudo-channel $\mathbf{G}$ and the true channel $\mathbf{H}$ is well conditioned. With the unitary assumption on $\mathbf{S}$, i.e., $\mathbf{S}\mathbf{S}^{\ast}=\mathbf{I}$, it follows from \eqref{eq:defn_G} that $\mathbf{H}=\mathbf{G}\mathbf{S}^{\ast}$. The channel estimate $\hat{\mathbf{H}}$ is then 
\begin{equation}
\label{eq:H_hat}
\hat{\mathbf{H}}=\hat{\mathbf{G}}\mathbf{S}^{\ast}.
\end{equation}
The choice of $\mathbf{S}$ within the unitary class is critical for the successful recovery of $\mathbf{G}$. We explain how $\mathbf{S}$ controls the channel estimation performance in Sec.~\ref{sec:training_design}.
\par The proposed algorithms solve for $\mathbf{G}$ instead of $\mathbf{H}$ to achieve a low complexity implementation of gradient ascent. To explain this argument, we define $\mathbf{W}\in \mathbb{C}^{N\times N}$ as the optimization variable corresponding to $\mathbf{H}$ in \eqref{eq:Yblk_N} and consider an $\mathbf{S}$ in $\mathbb{R}^{N \times N}$. The log-likelihood function of $\mathbf{W}$, defined as $\mathcal{C}_{\mathbf{Y}}(\mathbf{W})$, can be expressed as $\mathcal{C}_{\mathbf{Y}}(\mathbf{W})=\mathcal{L}_{\mathbf{Y}}(\mathbf{W}\mathbf{S})$. Gradient ascent algorithms that solve for $\mathbf{H}$ compute $\nabla \mathcal{C}_{\mathbf{Y}}(\mathbf{W})$ in every iteration. The gradient can also be expressed as $\nabla \mathcal{C}_{\mathbf{Y}}(\mathbf{W})=\nabla \mathcal{L}_{\mathbf{Y}}(\mathbf{WS})\mathbf{S}^T$. Now, computing $\nabla \mathcal{C}_{\mathbf{Y}}(\mathbf{W})$ requires two additional matrix multiplications when compared to $\nabla \mathcal{L}_{\mathbf{Y}}(\mathbf{X})$. The first matrix multiplication is due to $\mathbf{WS}$, and the second is between the gradient of $\mathcal{L}_{\mathbf{Y}}(\cdot)$ with $\mathbf{S}^T$. As the complexity of multiplying two $N \times N$ matrices is $\mathcal{O}(N^3)$, each gradient ascent step to maximize $\mathcal{C}_{\mathbf{Y}}(\mathbf{W})$ has $\mathcal{O}(N^3)$ higher complexity than its counterpart that maximizes $\mathcal{L}_{\mathbf{Y}}(\mathbf{X})$. 
\par At the end of gradient ascent iterations, algorithms that maximize $\mathcal{C}_{\mathbf{Y}}(\mathbf{W})$ and $\mathcal{L}_{\mathbf{Y}}(\mathbf{X})$ provide estimates of $\mathbf{H}$ and $\mathbf{G}$. For algorithms that estimate the pseudo-channel $\mathbf{G}$, the channel estimate is derived using  \eqref{eq:H_hat} for an additional complexity of $\mathcal{O}(N^3)$. As such a multiplication is performed only once, the proposed methods have a lower complexity than comparable gradient ascent methods that directly solve for $\mathbf{H}$ using the same number of iterations.
\vspace{-1mm}
\subsection{Training design for channel estimation}\label{sec:training_design}
\par The success of our channel estimation methods is determined by the underlying pseudo-channel optimization algorithms. It can be observed from \eqref{eq:defn_G} and \eqref{eq:H_hat} that the error in the channel estimate, i.e., $\Vert \mathbf{H}-\hat{\mathbf{H}}\Vert_{\mathrm{F}}$,  is $\Vert (\mathbf{G}-\hat{\mathbf{G}})\mathbf{S}^{\ast}\Vert_{\mathrm{F}}$ when $\mathbf{S}$ is unitary. Furthermore, $\Vert (\mathbf{G}-\hat{\mathbf{G}})\mathbf{S}^{\ast}\Vert_{\mathrm{F}}=\Vert \mathbf{G}-\hat{\mathbf{G}}\Vert_{\mathrm{F}}$ for a unitary $\mathbf{S}$. As $\Vert \mathbf{H}-\hat{\mathbf{H}}\Vert_{\mathrm{F}}=\Vert \mathbf{G}-\hat{\mathbf{G}}\Vert_{\mathrm{F}}$, reconstruction guarantees corresponding to \eqref{eq:generic_optim} can be used to study channel estimation performance. Prior work has shown that one-bit matrix recovery using \eqref{eq:generic_optim} achieves good performance when the energy in $\mathbf{G}$ is distributed across all its entries~\cite{davenport_one_bit}. To this end, the training block $\mathbf{S}$ must be chosen such that the maximum entry in $\mathbf{G}=\mathbf{HS}$ is small enough for any realistic channel $\mathbf{H}$. 
\par We use two different training solutions to study our algorithms. The first training solution defines $\mathbf{S}$ as a unitary discrete Fourier transform (DFT) matrix. Such a training is used in IEEE 802.11ad systems in which the TX performs DFT-based beam search~\cite{11ad}. The second training solution is defined by a unit norm Zadoff-Chu sequence $\mathbf{z}$~\cite{nitinglobecom}. In the ZC-based training, the $k^{\mathrm{th}}$ column of $\mathbf{S}$ is a $k$-circulant shift of $\mathbf{z}$. Although both the DFT-based and the ZC-based training blocks are unitary, the maximum entries of the corresponding pseudo-channels can differ significantly. To explain this difference, we consider an example of $\mathbf{H}=\mathbf{1}\mathbf{1}^T$, where $\mathbf{1}$ denotes a vector of ones. For such a channel, it can be verified that the maximum pseudo-channel entry is $\sqrt{N}$ with the DFT-based training and is $1$ with the ZC-based training. As the ZC-based training results in a lower pseudo-channel maximum, it is expected to achieve better one-bit channel reconstruction than the DFT-based training~\cite{davenport_one_bit}.
\par The channel estimation techniques discussed in Sec.~\ref{sec:algorithms} consider the special case of $\Np=N$. For $\Np<N$, the TX transmits $\Np$ columns of $\mathbf{S}$ at random. In this case, the received channel measurements is an $N\times \Np$ submatrix of $\mathbf{Y}$ in \eqref{eq:Yblk_N}. Our algorithms maximize the likelihood defined by the $\Np N$ channel measurements. For $\Np>N$, we define $\tilde{\mathbf{S}} \in \mathbb{C}^{N \times \Np}$ as the transmitted pilot block. A straightforward extension of our channel estimation technique is one that solves for $\tilde{\mathbf{G}}=\mathbf{H}\tilde{\mathbf{S}}$. Such an approach, however, requires optimization over $\Np N>N^2$ complex variables in $\tilde{\mathbf{G}}$ and can result in a high complexity. Gradient ascent techniques that directly solve for $\mathbf{H}$ also result in a high complexity as they require matrix multiplications with $\tilde{\mathbf{S}}$ and $\tilde{\mathbf{S}}^{\ast }$ in each gradient step. 
\par We propose a phase offset-based training solution that allows a low complexity extension of our algorithms for $\Np>N$. For simplicity, we assume $\Np$ to be an integer multiple of $N$ and define $B=\Np/N$. The phase offsets in our training are linearly spaced angles in $[0,\pi/2)$, i.e., $\theta_b=\pi b/(2 B)$ for $b \in \{0,1,\cdots B-1\}$. The proposed training is defined by $\tilde{\mathbf{S}}=[\mathbf{S}e^{\imj \theta_0},\mathbf{S}e^{\imj \theta_1},\cdots, \mathbf{S}e^{\imj \theta_{B-1}}]$. When the TX transmits $\tilde{\mathbf{S}}$, the RX receives one-bit quantized versions of $\{\mathbf{G}e^{\imj \theta_b}\}^{B-1}_{b=0}$. For $\Np>N$, our algorithms solve for the $N\times N$ pseudo-channel matrix $\mathbf{G}$. In this case, the likelihood of the variable corresponding to $\mathbf{G}$ is a sum of $B$ functions where each function is associated with an angle $\theta_b$. Each of these $B$ functions is separable in the complex variables $\{G_{k,\ell}\}^{N}_{k,\ell=1}$. The expression for $\mathcal{L}_{\mathbf{Y}}(\mathbf{X})$ is different from \eqref{eq:likelihood} because $Y_{k,\ell}^{\mathrm{R}}$ depends on both $G_{k,\ell}^{\mathrm{R}}$ and $G_{k,\ell}^{\mathrm{I}}$ when $\theta_b \neq 0$. Due to space constraints, we do not provide an explicit expression for $\mathcal{L}_{\mathbf{Y}}(\mathbf{X})$. The separable structure of the likelihood with the proposed training allows low complexity gradient computations when compared to the case with unstructured training. 
\section{Simulations}
\par We consider a mmWave MIMO system operating at a carrier frequency of $28\, \mathrm{GHz}$. Both the TX and the RX are equipped with a half-wavelength spaced uniform linear array of $N=16$ antennas. We evaluate our algorithms using $100$ urban micro non-line-of-sight channels from the NYU simulator~\cite{NYUSIM}, for a TX-RX separation of $15\, \mathrm{m}$. The channels are scaled so that their Frobenius norm is $N$. The SNR of the channel measurements is defined as $\mathrm{SNR} = 10\,\mathrm{log}_{10}(1/\sigma^2)$. The proposed methods require tuning the parameter $\beta$. We used $\beta=20.1$ as $90 \%$ of the channels had a nuclear norm less than $20.1$. The maximum number of iterations is set to $T_{\mathrm{max}}=80$. For both the algorithms, the stopping criterion is defined by setting $\epsilon=10^{-10}$. A step size of $\eta = 0.1/ B$ is used for the PGA algorithm. The step size is scaled down by $0.5$ whenever the likelihood decreases at any iteration. In our simulations, we observed large entries in the gradient, i.e., $\nabla \mathcal{L}_{\mathbf{Y}}(\mathbf{X})$, at high SNRs. Such high magnitude gradients can prevent the algorithms from convergence. To overcome this problem, the standard deviation $\sigma$ in the likelihood is replaced by $\sigma_{\mathrm{likel}}=\mathrm{max}(0.5, \sigma)$. It is important to note that such a clipping is only performed for numerical stability and the received measurements are still acquired under a noise variance of $2\sigma^2$. An implementation of the proposed algorithms for the phase offset-based training can be found on our page~\cite{Code_one_bit}.
\par We evaluate the one-bit compressed sensing (CS) method in~\cite{mo2016channel} and the maximum likelihood method in~\cite{junil_one_bit_chest} for benchmarks. A 2D-DFT is used as the sparsifying dictionary for channel estimation with CS. The CS method in \cite{mo2016channel} uses a message passing algorithm called EM-BG-GAMP \cite{EMGAMP}. For the maximum likelihood technique in~\cite{junil_one_bit_chest}, we use the Frobenius norm constraint, i.e., $\Vert \mathbf{H} \Vert_{\mathrm{F}} \leq 16$. Such a constraint was used in~\cite{junil_one_bit_chest} to avoid an unbounded solution. The ZC-based and the DFT-based training matrices defined in Sec.~\ref{sec:training_design} are used to obtain channel measurements. We evaluate the algorithms in terms of the normalized mean squared error (NMSE) and the achievable rate. We define the NMSE as the average of $\Vert \mathbf{H}- \kappa \hat{\mathbf{H}} \Vert^2_{\mathrm{F}}/\Vert \mathbf{H}\Vert^2_{\mathrm{F}}$ over several channel realizations, where $\kappa = \mathrm{arg\,min} \Vert \mathbf{H}-\kappa \hat{\mathbf{H}}\Vert_{\mathrm{F}}$ for a given $\mathbf{H}$ and $ \hat{\mathbf{H}}$. The rate achieved with the channel estimates is computed using the expression in \cite{mo2016channel}. The training overhead for channel estimation is ignored in computing the achievable rate. 
\begin{figure}[h!]
\centering
\includegraphics[trim=1.5cm 6.75cm 2.5cm 7.5cm,clip=true,width=0.37 \textwidth]{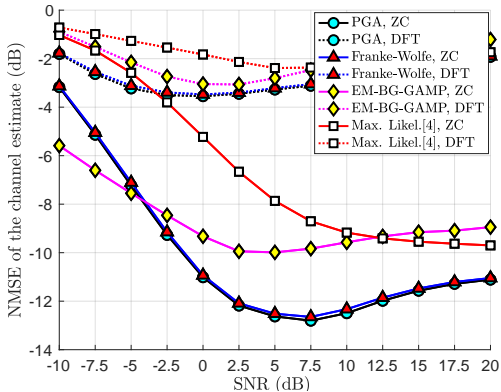}
  \caption{The plot shows the NMSE of the channel estimate with SNR for $\Np=64$ pilot transmissions. The increase in NMSE at high SNRs is due to stochastic resonance~\cite{stochastic_res}.}
  \label{fig:NMSE_vs_SNR}
\end{figure}
\begin{figure}[h!]
\centering
\includegraphics[trim=1.5cm 6.5cm 2.5cm 7.5cm,clip=true,width=0.37 \textwidth]{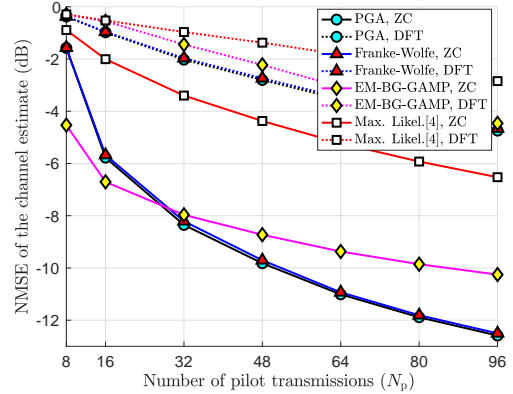}
    \vspace{-1mm}
  \caption{The plot shows NMSE with the number of pilots for an SNR of $0\, \mathrm{dB}$. The DFT-based training results in poor performance due to large variation in the corresponding pseudo-channel entries.}
  \label{fig:NMSE_vs_pilots}
      \vspace{-1mm}
\end{figure}
\begin{figure}[h!]
\centering
\includegraphics[trim=1.5cm 6.5cm 2.5cm 7.75cm,clip=true,width=0.37 \textwidth]{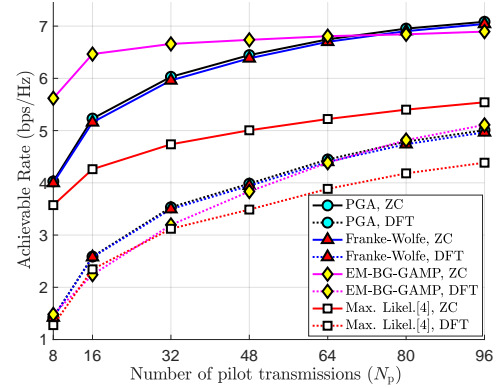}
  \caption{The plot shows the achievable rate with the number of pilots for an SNR of $0\, \mathrm{dB}$. The results indicate that $\Np=48$ shifted ZC-based pilot transmissions are sufficient to achieve a reasonable rate.}
  \label{fig:Rate_vs_pilots}
  \vspace{-3mm}
\end{figure}
\par From Fig.~\ref{fig:NMSE_vs_SNR}, it can be observed that the proposed one-bit channel estimation methods outperform techniques based on maximum likelihood and one-bit CS over a wide range of SNR. The proposed methods learn dictionaries that result in a low-rank channel representation when compared to one-bit CS techniques that use the 2D-DFT. The poor performance of our algorithms for $\mathrm{SNR}<-5\,\mathrm{dB}$ can be attributed to errors in learning the dictionary. The maximum likelihood method in~\cite{junil_one_bit_chest} performs poor as it does not exploit the low-rank structure of channel matrices. It can be observed from Fig.~\ref{fig:NMSE_vs_pilots} and Fig.~\ref{fig:Rate_vs_pilots} that our gradient ascent-based algorithms result in better channel estimates for $\Np>48$ pilots, at an SNR of $0\, \mathrm{dB}$. For the channel realizations in our simulations, we observed that the mean peak-to-average magnitude ratio of the entries in $\mathbf{G}$ was $3.99\, \mathrm{dB}$ and $16.74\, \mathrm{dB}$ for the ZC- and the DFT-based training blocks. The less ``spiky'' pseudo-channel structure with the ZC-based training results in better one-bit channel estimation performance, as shown in Fig.~\ref{fig:NMSE_vs_pilots}.
 \section{Conclusions}
We have developed gradient ascent-based algorithms for low-rank channel estimation from one-bit measurements. We have also proposed a phase offset-based training to reduce the complexity of our algorithms. Our results indicate that algorithms which exploit the low-rank structure in mmWave channels can achieve better channel reconstruction than those that exploit sparsity. Our results also show that the use of Zadoff-Chu-based training results in better channel estimates than the standard DFT-based training in one-bit receivers. 
\bibliographystyle{IEEEtran}
\bibliography{refs2}
\end{document}